\documentclass[]{spie}  
\usepackage[]{graphicx}
\usepackage{amsmath,amssymb}
\title{Observation of bosonic coalescence and fermionic anti-coalescence with indistinguishable photons}

\author{Guillaume Adenier, Joakim Bergli, Andreas P. Th\"{o}rn and Arnt Inge Vistnes
\skiplinehalf
Department of Physics, University of Oslo, Sem S{\ae}lands vei 24, 0316 Oslo, Norway
}

\begin{document}
  \maketitle

\begin{abstract}
The symmetrization postulate asserts that the state of particular species of particles can only be of one permutation symmetry type: symmetric for bosons and antisymmetric for fermions. We report some experimental results showing that pairs of photons indistinguishable by all degrees of freedom can exhibit not only a bosonic behavior, as expected for photons, but also a surprisingly sharp fermionic behavior under specific conditions.
\end{abstract}



\section{Introduction}

The symmetrization postulate can be expressed very succinctly : ``states containing several identical elementary particles are, according to the species, either symmetric (bosons) or antisymmetric (fermions)''\cite{Messiah}. It restricts quite strongly the Hilbert space accessible to two-boson systems since their full quantum state must remain symmetrical \cite{Zeilinger94,kwiat97,Parks}, and has some important application in the description of interference phenomena that are typically quantum mechanical in nature, such as the Hong-Ou-Mandel effect \cite{HOM}.

Antisymmetric states in a particular degree of freedom---such as the polarization degree of freedom---have nevertheless long been observed (see for instance \cite{Zeilinger98,Bouwmeester,Michler,Monken,DeMartini05}), but this is usually achieved by using photons that are distinguishable by their spatial mode. Bosonic symmetry imposes that photons with antisymmetric polarization state vector impinging on two different input of a 50/50 beamsplitter end up in different outputs at the beam splitter. The antisymmetry of the polarization degree of freedom is then compensated by the antisymmetry of the spatial degree of freedom, so that the overall state remains symmetric \cite{Zeilinger94}. This idea is at the heart of Bell-state analysis \cite{kwiat97,kwiat98,kwiat07}, and has many applications such as quantum teleportation, quantum dense-coding \cite{Bouwmeester} or quantum random walks \cite{Sansoni}.

The results that we would like to report here is different because in our experimental setup the photons are collinear and incident on the \emph{same} input port $|\mathrm{c}\rangle$ of the beam-splitter. The spatial modes of the two orthogonally polarized photons are not distinguished in our experimental setup, so that it is a priori the same for both photons, and therefore symmetric by particle exchange. Under these circumstances, one would expect in particular that when the photons are made indistinguishable by all degrees of freedom, they would behave exclusively like bosons. As we will see, the pairs of photons can nevertheless display not only a bosonic behavior, as expected for photons, but also in some cases exhibit a behavior that one would rather associate with fermions. We will report here on the experimental findings, and sketch how the symmetry can possibly be recovered, with the use of an additional, but presently unknown, degree of freedom.

\section{Experimental setup}\label{Expsetup}

Our source of polarization-entangled photons is directly inspired by the setup implemented by Kuklewicz et al.\cite{Kuklewicz}. The pairs of photons are obtained by type-II spontaneous parametric down conversion (SPDC) in a periodically-poled crystal of potassium-titanyl-phosphate (ppKTP), under quasi-phase matching (QPM) condition. The pump, a continuous-wave laser at 405 nm, and the down converted photons are all collinear. The temperature of the ppKTP crystal is controlled by thermoelectric Peltier temperature controller.

\begin{figure}
\begin{center}
\includegraphics[width=12cm]{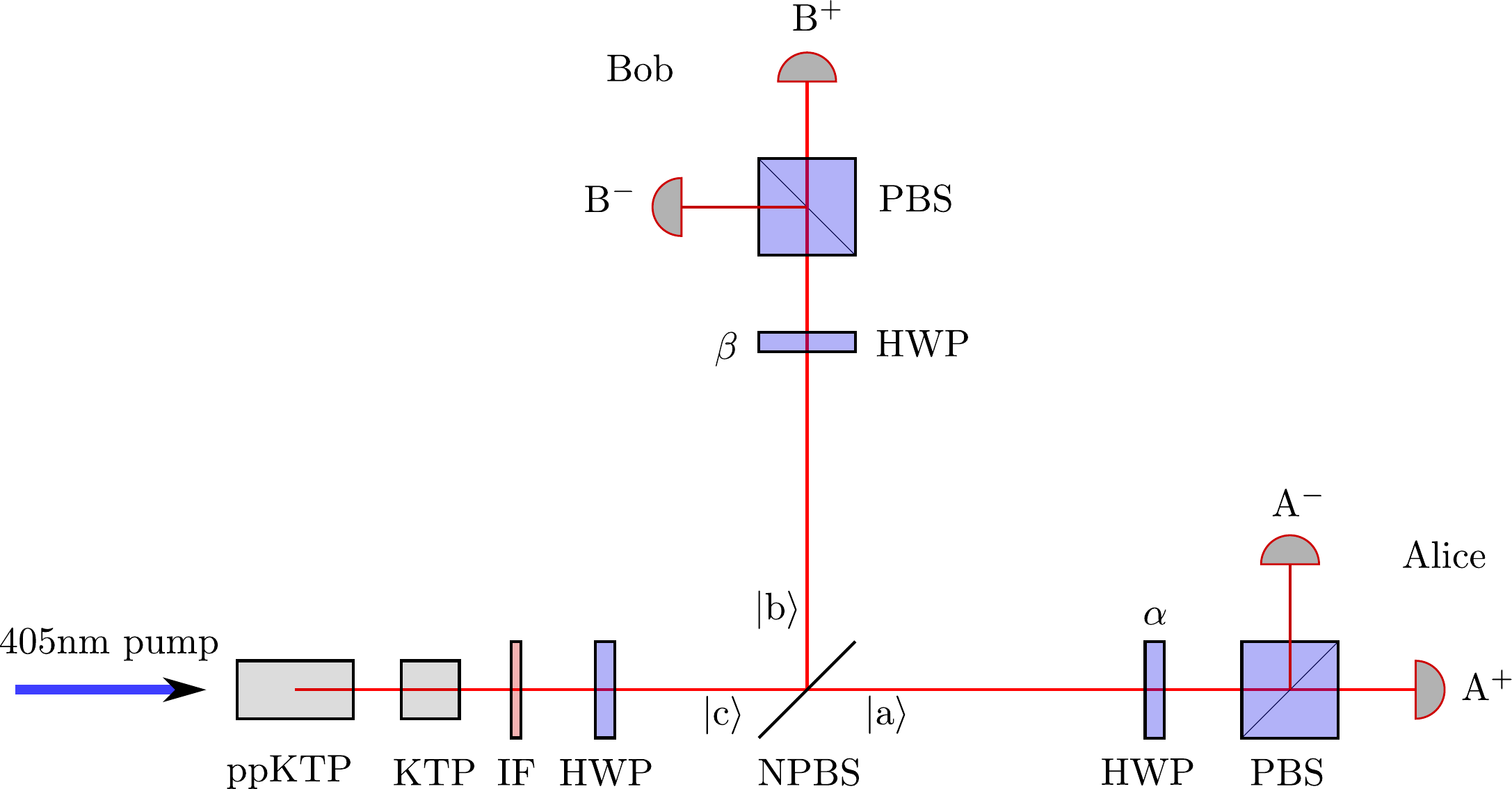}
\end{center}
\caption{\label{RTfig} (Color online) Experimental setup. The pairs of orthogonally polarized photons are down converted from the 405 nm pump in the ppKTP crystal. They are time compensated (in a KTP crystal), frequency filtered (IF), and spatially filter (Iris) before being dispatched to Alice and Bob by a non-polarizing beam splitter (NPBS). At each measuring station, a half-wave plate (HWP) rotates the polarization of the photons, and a PBS projects them to a fixed basis. The output of the PBS feed four detectors, $\mathrm{A}^+$ and $\mathrm{A}^-$ for Alice, $\mathrm{B}^+$ and $\mathrm{B}^-$ for Bob.}
\end{figure}

After filtering out the pump with a series of dichroic and interference filters (in particular one with a bandwidth of 1 nm), the collinear down converted photons are dispatched by a 50/50 non-polarizing beam splitter (NPBS) to two polarization analyzers. We label the polarization analyzer in the transmitted beam as ``Alice'', and the one in the reflected beam as ``Bob''. Each consists of a half-wave plate, $\alpha$ for Alice and $\beta$ for Bob, that rotates the polarization of the field, followed by a polarizing beam splitter (PBS) that projects it in a fixed basis $\{|H\rangle,|V\rangle\}$.

The outputs of each PBS are feeding two detectors, labeled  $\mathrm{A}^+$ and $\mathrm{A}^-$ for Alice, and $\mathrm{B}^+$ and $\mathrm{B}^-$ for Bob (see Fig.~\ref{RTfig}). The detectors are four avalanche photodiodes SPCM-AQRH-16, from Perkin-Elmer, with a detection efficiency specified at 60\% by the manufacturer with 25 dark counts per seconds. The detection events (clicks) are time-tagged with nominal picosecond precision by a Multichannel Picosecond Event Timer (Hydraharp 400, from Picoquant), and saved to disk for on-the-fly analysis (when the data flow is not too important), as well as subsequent analysis.

No detected events are discarded, so that the coincidence analysis can be performed after-the-fact with adjustable parameters (size of the coincidence window and timing-offset). The flexibility of this acquisition setup based on the recording of the detection time of the photons is of course largely inspired by the landmark Innsbruck experiment performed by Gregor Weihs et al. \cite{Weihs}. Having a complete record of all the detected events allows us to be thorough in the coincidence analysis: the rate of single counts of course, but also all types of rate of coincidences.

A measured rate of coincidences between two specific detectors denotes the number of occurrences of a simultaneous click (i.e., to within about an arbitrary time window of a few nanoseconds) of these two detectors, during an acquisition time of exactly 1 second. The four possible rates of coincidences between Alice's and Bob's detectors are denoted $R_\mathrm{ab}^{++}$, $R_\mathrm{ab}^{+-}$, $R_\mathrm{ab}^{-+}$ and $R_\mathrm{ab}^{--}$, where the first superscript index indicates which of Alice's detector is considered, and the second which of Bob's detector is. We call \emph{double-counts} the coincidences between two detectors located in the same measuring station (Alice or Bob), and we label them $R^{\pm}_\mathrm{aa}$ for Alice's double-counts, and $R^{\pm}_\mathrm{bb}$ for Bob's. As we will see, these rate of double-counts---which are usually not recorded---provide some useful information about the exact state of the pairs of photons.

\section{Polarization entanglement}\label{theory}

\subsection{Spontaneous parametric down conversion}

The theoretical description to obtain polarization entanglement from collinear type-II spontaneous parametric down conversion (SPDC) is rather simple (see for instance \cite{Rubin94,TSuhara,TSuhara2007,Martin08,Martin10,KuklewiczPhD,Sciarrino}). In the nonlinear crystal, the interaction Hamiltonian is \cite{Rubin94}
\begin{equation}\label{Hamilton }
    \hat{H}=\epsilon_0\int_V \mathop{dr^3} \chi^{(2)}\hat{E}^+_\mathrm{p}\hat{E}^-_\mathrm{s}\hat{E}^-_\mathrm{i} \;+ \; h.c,
\end{equation}
where the indices stand respectively for \emph{pump}, \emph{signal} and \emph{idler}. A photon from the pump can be spontaneously down-converted to two daughter photons.
In a periodically poled crystal, this process must fulfill the quasi phase-matching (QPM) conditions, for the angular frequencies
\begin{equation}\label{omegaSPDC}
    \omega_\mathrm{p}=\omega_\mathrm{s}+\omega_\mathrm{i},
\end{equation}
and for the wave numbers
\begin{equation}\label{kSPDC}
    k_\mathrm{p}=k_\mathrm{s}+k_\mathrm{i}+\frac{2\pi}{\Lambda},
\end{equation}
where $\Lambda$ is the period of the poling in the ppKTP crystal.

For collinear SPDC confined to a single spatial mode, the calculation to first order perturbation theory for the quantum state of the pairs of down-converted photons at the output of the nonlinear crystal is \cite{Teich02,Nam03}
\begin{equation}\label{SPDC}
  |\psi\rangle \propto \int d\omega \;\tilde{\Phi}(\omega)\; \hat{a}_\mathrm{s}^\dag(\frac{\omega}{2}+\omega)\;
  \hat{a}_\mathrm{i}^\dag(\frac{\omega}{2}-\omega)|0\rangle,
\end{equation}
where $\hat{a}_\mathrm{s}^\dag(\omega)$ and $\hat{a}_\mathrm{i}^\dag(\omega)$ are the creation operators for signal and idler with orthogonal polarization in frequency mode $\omega$, and where the integral is taken from $-\infty$ to $+\infty$. The state function $\tilde{\Phi}(\omega)$ is normalized as  $\int d\omega\;\tilde{\Phi}(\omega)=1$
and depends on the physical structure of the nonlinear crystal.

Since we are also selecting only the pairs that have a wavelength of 810 nm (with an interference filter with~1~nm bandwidth in our case), it can be simplified to a single-mode pair description \cite{Rubin94,TSuhara,TSuhara2007}
\begin{equation}\label{HVstate}
    |\psi\rangle \propto \hat{a}_\mathrm{H}^\dag\; \hat{a}_\mathrm{V}^\dag |0\rangle=|\mathrm{H}\rangle|\mathrm{V}\rangle,
\end{equation}
where we have substituted the idler and signal notation for the horizontal (H) and vertical (V) polarizations relevant to our laboratory setup.

\subsection{Postselection and entanglement}

In order to dispatch the photons to Alice and Bob, the pairs of collinear photons produced in the ppKTP crystal are sent to a non polarizing beam splitter (NPBS). For a photon impinging with a spatial mode $|\mathrm{c}\rangle$ on an ideal 50/50 NPBS, the output state in terms of the transmitted mode $|\mathrm{a}\rangle$ and the reflected mode $|\mathrm{b}\rangle$ (sent respectively to Alice and Bob; see Fig.~\ref{RTfig}) depends on the initial polarization of the photon \cite{TSuhara}:
\begin{equation}\label{BStransfo}
\begin{aligned}
    |\mathrm{H}\rangle|\mathrm{c}\rangle &   \xrightarrow{\mathrm{NPBS}} \frac{1}{\sqrt{2}} \big(|\mathrm{H}\rangle|\mathrm{a}\rangle+i|\mathrm{H}\rangle|\mathrm{b}\rangle \big)
    \\
     |\mathrm{V}\rangle|\mathrm{c}\rangle &   \xrightarrow{\mathrm{NPBS}} \frac{1}{\sqrt{2}} \big(|\mathrm{V}\rangle|\mathrm{a}\rangle-i|\mathrm{V}\rangle|\mathrm{b}\rangle \big)
\end{aligned}
\end{equation}

If we now consider two orthogonally polarized photons impinging on the beam splitter with the same input spatial mode, we can write, using the spatial mode as a shorthand index for the polarization mode:
\begin{equation}\label{HVBS}\nonumber
    |\mathrm{H}\rangle_\mathrm{c}|\mathrm{V}\rangle_\mathrm{c} \xrightarrow{\mathrm{NPBS}}
\frac{1}{2} \big(
    |\mathrm{H}\rangle_\mathrm{a}|\mathrm{V}\rangle_\mathrm{a}
    -i|\mathrm{H}\rangle_\mathrm{a}|\mathrm{V}\rangle_\mathrm{b}
+i|\mathrm{H}\rangle_\mathrm{b}|\mathrm{V}\rangle_\mathrm{a}
    +|\mathrm{H}\rangle_\mathrm{b}|\mathrm{V}\rangle_\mathrm{b}
    \big).
\end{equation}
The usual argument at this point \cite{Rubin94,TSuhara,TSuhara2007,Martin08,Martin10,KuklewiczPhD,Sciarrino}) is that the cases in which the two photons exit through the same port (that is, $|\mathrm{H}\rangle_\mathrm{a}|\mathrm{V}\rangle_\mathrm{a}$ and $|\mathrm{H}\rangle_\mathrm{b}|\mathrm{V}\rangle_\mathrm{b}$) can be discarded because of the postselection of the photons. Only those pairs with one photon for Alice and one photon for Bob are considered of interest.

After making the substitution $|\mathrm{H}\rangle_\mathrm{b}|\mathrm{V}\rangle_\mathrm{a}\rightarrow|\mathrm{V}\rangle_\mathrm{a}|\mathrm{H}\rangle_\mathrm{b}$,
and renormalizing, the state of the pairs of photons detected in coincidence by Alice and Bob can be written as the singlet state:
\begin{equation}\label{singlet}
    |\Psi^-\rangle_\mathrm{ab} = \frac{1}{\sqrt{2}} \big[|\mathrm{H}\rangle_\mathrm{a}|\mathrm{V}\rangle_\mathrm{b}-|\mathrm{V}\rangle_\mathrm{a}|\mathrm{H}\rangle_\mathrm{b} \big],
\end{equation}
which is a polarization-entangled state.

Starting from this post-selected state, the rates of coincidences by Alice and Bob take the simple and well-known form associated to the singlet state:
\begin{equation}\label{expoptC}
\begin{aligned}
    R^{++}_\mathrm{ab} \approx R^{--}_\mathrm{ab} &\varpropto \frac{1}{2} \sin^2 2(\alpha-\beta)\\
    R^{+-}_\mathrm{ab} \approx R^{-+}_\mathrm{ab} &\varpropto \frac{1}{2} \cos^2 2(\alpha-\beta).
\end{aligned}
\end{equation}

\begin{figure}
\begin{center}
a)\includegraphics[width=8.10cm]{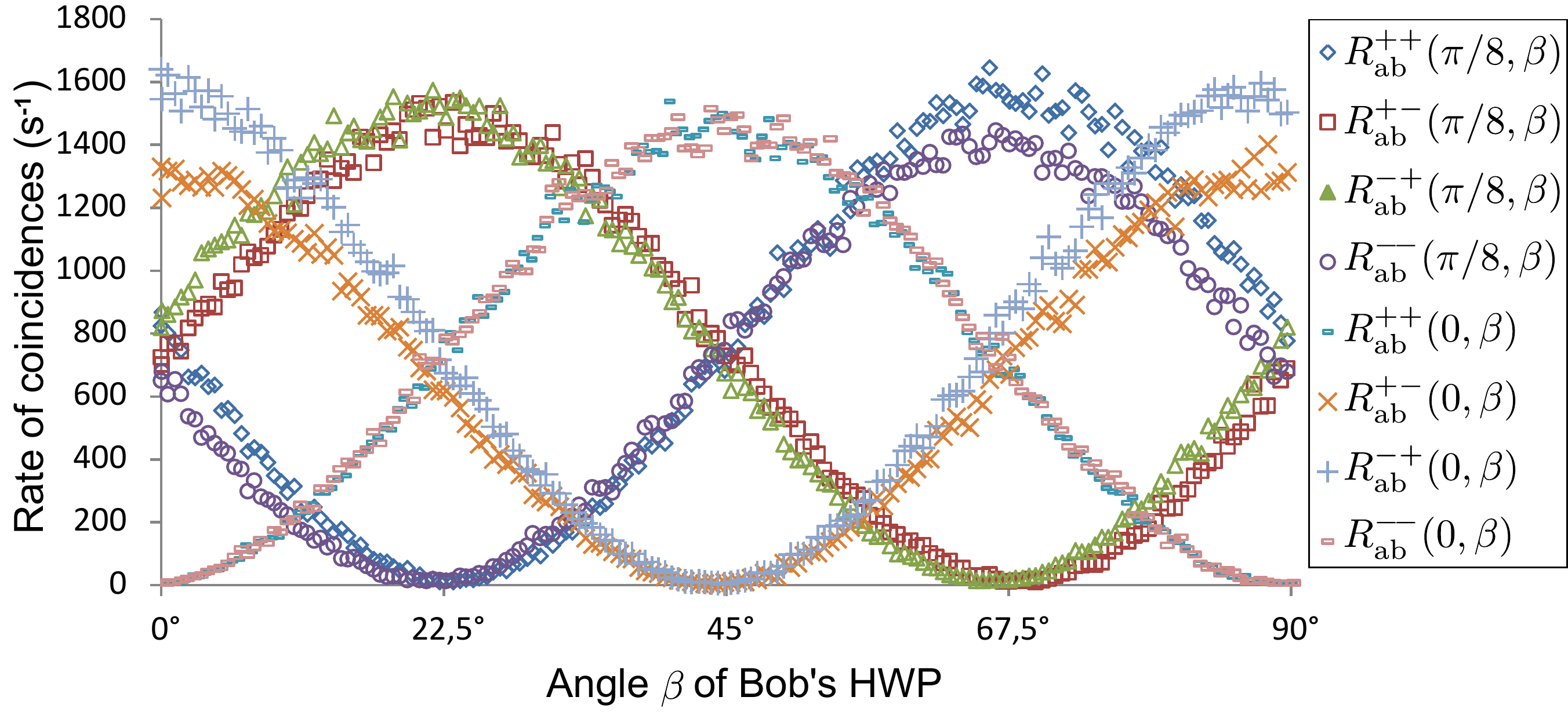}
b)\includegraphics[width=8.10cm]{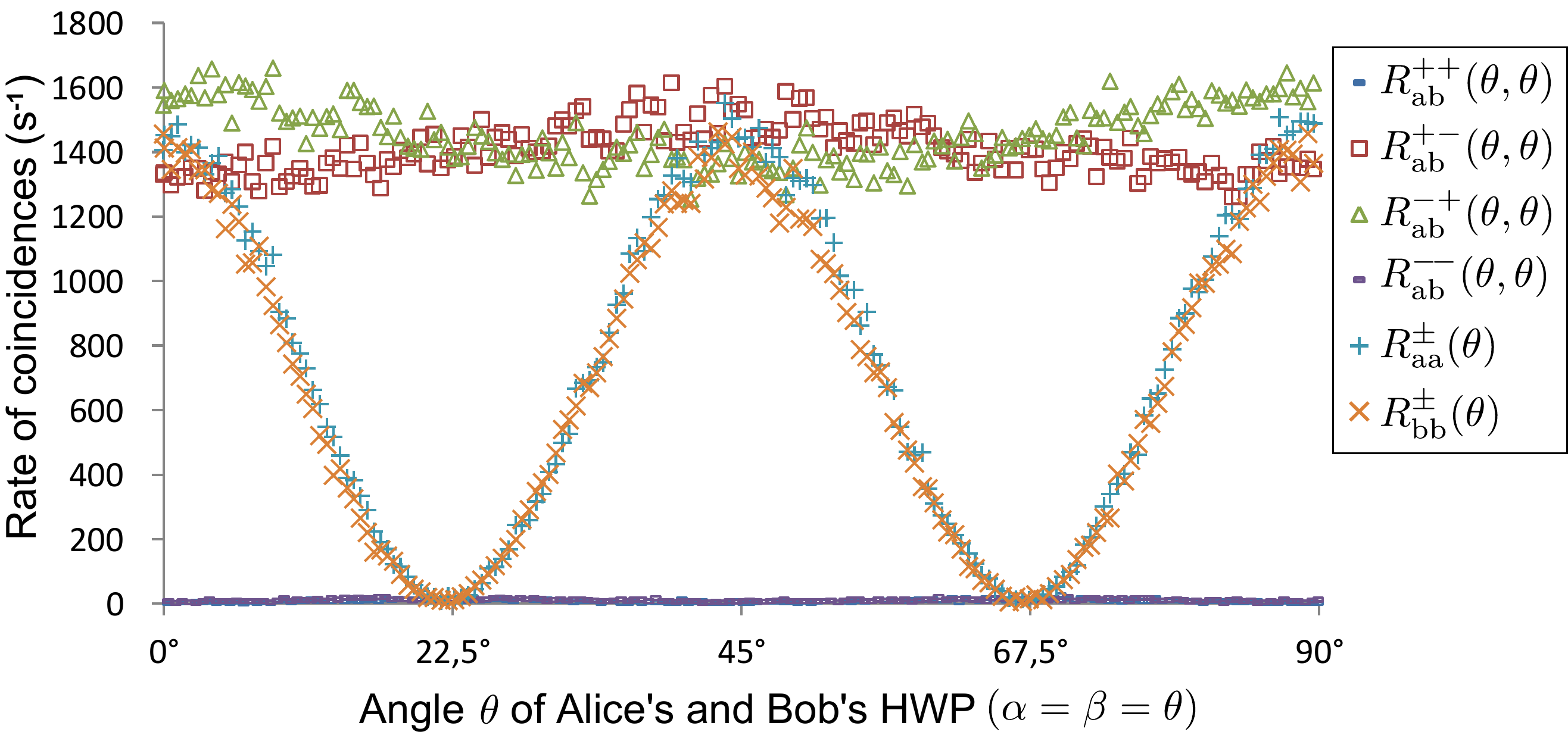}
\end{center}
\caption{\label{FigOpt} (Color online) a) Rates of coincidences in fixed bases while the temperature of the crystal is maintained at the optimum temperature of $35.1^\circ$C. Alice keeps her measurement setting fixed, at $\alpha=0$ (rectilinear basis) or $\alpha=\pi/8$ (diagonal basis), while Bob's HWP setting $\beta$ is varied from 0 to 90. The coincidences exhibit a visibility of 99.6\% in the rectilinear basis and of 98.5\% in the diagonal basis (without substraction of accidental), which would amount to a CHSH function $S\simeq2.80$. b) Rates of coincidences and double-counts during a \emph{twin scan} while the temperature of the crystal is maintained at the optimum temperature of $35.1^\circ$C. By contrast, the double-counts exhibit a strong rotational dependence on $\theta$.}
\end{figure}

With good approximation---the visibility (or contrast) of the coincidences is slightly less than ideal, with a visibility of 99.6\% in the rectilinear basis and of 98.5\% in the diagonal basis (without substraction of accidental)---the rate of coincidences that we measure are indeed of this form (see Fig.~\ref{FigOpt}-a).

\subsection{Double-counts}

Thanks to the flexibility of our acquisition setup, since we are recording the time of all detection events, there is no reason for us to discard the cases when the two photons exit through the same port $|\mathrm{H}\rangle_\mathrm{a}|\mathrm{V}\rangle_\mathrm{a}$ or $|\mathrm{H}\rangle_\mathrm{b}|\mathrm{V}\rangle_\mathrm{b}$). The coincidences that these photons can generate are the double-counts, that is, coincidences between the two detectors located in the same output arm of the NPBS. As we will see, they can provide some useful information about the nature of the state associated with the pairs of photons.

Experimentally, the behavior of the double-counts can be seen comprehensively during what we call a ``twin scan'', that is, a measurement run during which the angles of the HWPs controlled by Alice and Bob are varied from 0 to 90, but kept equal at all time with $\alpha=\beta=\theta$. The rate of double counts that we have measured together with the coincidences during such a twin scan can be seen in Fig.~\ref{FigOpt}-b, and take the simple form
\begin{equation}\label{expoptD}
\begin{aligned}
    R^{\pm}_\mathrm{aa}&\varpropto \frac{1}{2} \cos^2 4\alpha\\
    R^{\pm}_\mathrm{bb}&\varpropto \frac{1}{2} \cos^2 4\beta,
\end{aligned}
\end{equation}
up to a less than perfect visibility of about 98\%.

The striking feature of the double-counts is that their rates drop close to zero in the diagonal bases, when $\alpha=\beta=\pi/8$, for both Alice and Bob.
In order to understand this behavior, we consider the case when both photons end up in Alice's location. Starting from the same state of Eq.~(\ref{HVBS}) that allowed us to derive the coincidences after the NPBS, we are thus post-selecting on the cases when both particles are detected by Alice, that is, with a spatial mode ``$\mathrm{a}$''. The state for these post-selected photons is:
\begin{equation}\label{doubleA}
    |\psi\rangle_\mathrm{aa} =|\mathrm{H}\rangle_\mathrm{a}|\mathrm{V}\rangle_\mathrm{a},
\end{equation}
which is in fact the same type of state as the single-mode pair state of Eq.~(\ref{HVstate}) before the NPBS, only this time we have written the spatial mode explicitly (i.e., Alice). The effect of Alice's half-wave plate oriented along $\alpha$ on the state of Eq.~(\ref{doubleA}) is
\begin{equation}\label{HVHWP}
\begin{aligned}
    |\mathrm{H}\rangle_\mathrm{a}|\mathrm{V}\rangle_\mathrm{a}
     \xrightarrow{\mathrm{HWP}\;\alpha} -\cos2\alpha\sin2\alpha&|\mathrm{H}\rangle_\mathrm{a}|\mathrm{H}\rangle_\mathrm{a}\\
    +\cos^2 2\alpha&|\mathrm{H}\rangle_\mathrm{a}|\mathrm{V}\rangle_\mathrm{a}\\
    -\sin^2 2\alpha&|\mathrm{V}\rangle_\mathrm{a}|\mathrm{H}\rangle_\mathrm{a}\\
    \cos2\alpha\sin2\alpha&|\mathrm{V}\rangle_\mathrm{a}|\mathrm{V}\rangle_\mathrm{a}.
\end{aligned}
\end{equation}

If we were dealing with distinguishable particles, we would be able to distinguish the cases $|\mathrm{H}\rangle_\mathrm{a}|\mathrm{V}\rangle_\mathrm{a}$ and $|\mathrm{V}\rangle_\mathrm{a}|\mathrm{H}\rangle_\mathrm{a}$, and we would therefore sum the probabilities associated with these two final states that can lead to a double count:
\begin{equation}\label{pdoubleAAdis}
    P^\mathrm{+-}_\mathrm{aa}(\alpha)+P^\mathrm{-+}_\mathrm{aa}(\alpha)=\cos^4 2\alpha+\sin^4 2\alpha=\frac{1}{2}(1+\cos^2 4\alpha).
\end{equation}
One should not forget however that in order to obtain the high quality of entanglement seen in Fig.~\ref{FigOpt}-a, the photons had to be made indistinguishable by all degrees of freedom except polarization. In particular, we have restricted their spatial modes with a small aperture iris, their wavelength with a 1 nm interference filter, and even compensate for the birefringence walk-off---that would have allowed to distinguish them by their arrival time---with the use of a compensating KTP crystal of half the length of the ppKTP crystal\cite{Shih,KuklewiczPhD,Kuklewicz}.

The final states $|\mathrm{H}\rangle_\mathrm{a}|\mathrm{V}\rangle_\mathrm{a}$ and $|\mathrm{V}\rangle_\mathrm{a}|\mathrm{H}\rangle_\mathrm{a}$ leading to double-counts are therefore experimentally indistinguishable, so that one should rather sum their probability amplitudes first, and then only square the modulus to get the probability \cite{Cohen}. The probability that two photons reaching together Alice's PBS would generate a double-count then takes the form
\begin{equation}\label{pdoubleAAident}
P^{\pm}_\mathrm{aa}(\alpha)=(\cos^2 2\alpha-\sin^2 2\alpha)^2=\cos^2 4\alpha,
\end{equation}
and similarly for Bob, which is exactly what is observed experimentally.

The absence of double-counts in the diagonal basis is interesting. Considering that the single counts (not shown here) are close to being rotationally invariant, it is safe to infer that the photons are then always exiting the PBS through the same output port. We cannot measure this effect directly, as our single-photon avalanche photodiodes cannot distinguish a single-photon absorption from a 2-photon absorption. Nevertheless, this effect has actually already been demonstrated in a similar configuration with photon-number resolving (PNR) detectors by Di Giuseppe \emph{et al.} \cite{Nam03}. They have observed that the probability $P(2,0)$ to detect two photons in a PNR detector located in the transmitted channel of Alice's polarizer---and similarly for Bob with a probability denoted $P(0,2)$---increases when the photons are gradually made indistinguishable by all degrees of freedom, including polarization. Their measurement is performed in fixed diagonal bases, for which the $|\mathrm{H}\rangle$ and $|\mathrm{V}\rangle$ photons are indistinguishable by their polarization, and with adjustable time delay $\tau$. In our case it is the opposite but the interpretation should be the same: the rate of double-counts drops when the photons are made more and more indistinguishable because the photons being bosons, they coalesce when they are made fully indistinguishable: two indistinguishable photons incident on such a polarization analog of the Hong-Ou-Mandel interferometer stick together as they exit the beam-splitter ports \cite{Nam03}.

As we are about to see, the above interpretation in terms of bosonic coalescence is insufficient since the exact opposite behavior---i.e., indistinguishable photons anti-coalescing as fermions would do---can be observed with the same experimental setup, at non-optimal temperatures of the ppKTP crystal.

\section{Non-optimal temperature of the ppKTP crystal}\label{nonoptimum}

So far, all the results have been obtained at optimum temperature of the ppKTP crystal (in our case $35.1^{\circ}\mathrm{C}$), for which the number of produced polarization entangled pairs is maximal. We now study the properties of our entangled photons when the temperature of the ppKTP crystal is brought away from this optimal temperature.
\subsection{Experimental observations}

\begin{figure}
\begin{center}
a)\includegraphics[width=8.1cm]{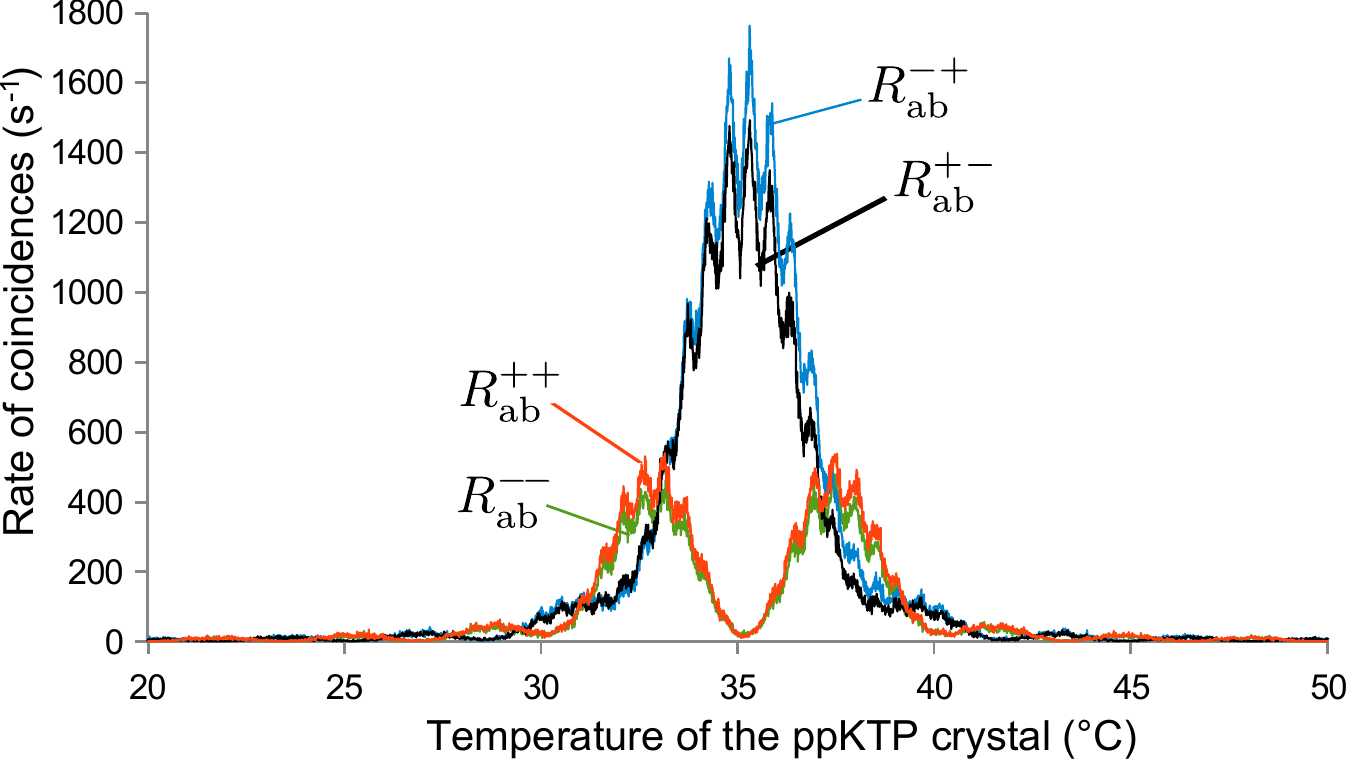}
b)\includegraphics[width=8.1cm]{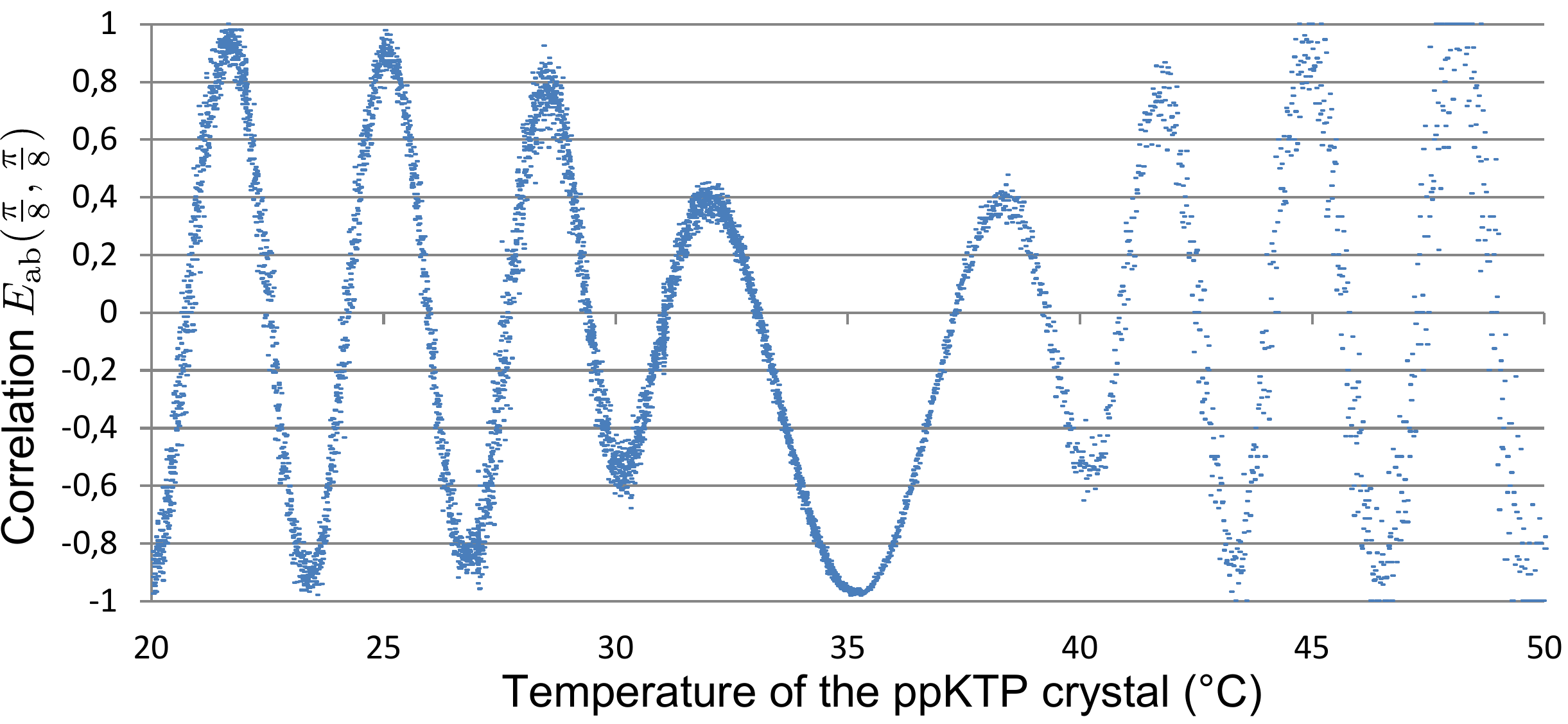}
\end{center}
\caption{\label{VaryTemp} (Color online) a) Rates of coincidences in the diagonal bases ($\alpha=\beta=\pi/8$) as a function of temperature of the ppKTP crystal. The rate of coincidences drops quickly away from the optimal temperature because the distribution of the wavelengths of the idler and signal are no longer centered on the 810 nm of our narrow bandwidth interference filter. b) Correlation in the diagonal bases ($\alpha=\beta=\pi/8$) as a function of temperature of the ppKTP crystal. Although the rate of coincidences drops quickly away from the optimum temperature of $35.1^\circ$C, the correlation does not in any way disappear, but oscillates with increasing amplitudes. The occurrences of positive correlation close to 1, indicating that the photons share the same polarization in the diagonal basis, is particularly worth investigating, given that we are operating in SPDC type II conditions, i.e., with orthogonally polarized photons.}
\end{figure}

The first noticeable effect of bringing the temperature of the ppKTP crystal away from the optimal temperature is that the rate of detected singles and coincidences quickly drops, as can be seen for the coincidences in Fig.~\ref{VaryTemp}-a. The reason for this behavior is that the spectrum of the down-converted photons depends on the temperature of the ppKTP crystal. Indeed, all the terms in the quasi-phase matching condition of Eq.~(\ref{kSPDC}) depend on the temperature of the ppKTP crystal \cite{Emanueli}. Now, because of the interference filter with a narrow bandwidth of 1 nm in our setup (see Fig.~\ref{RTfig}), we are nevertheless selecting those pairs of photons that happen to have the \emph{same} wavelength of 810 nm. So, as the temperature is brought away from the optimum temperature, the pairs of photons that match this strict wavelength criterion are less and less frequent, so that the number of coincidences drops quickly\footnote{We have observed that with an interference filter with wider bandwidth (10 nm) centered on 810 nm wavelength, the rate of single counts and coincidence counts is almost unaffected by the same range of temperature variation, which indicates that the wavelength of the down converted photons remain close to 810 nm, to within about 10 nm.}.

Even though the number of pairs that passes through diminishes away from the optimal temperature, we can still measure their coincidences, and their correlation. The correlation being defined as a linear combination of the coincidence rates, normalized by the sum of coincidences rates
\begin{equation}\label{correlation}
E_\mathrm{ab} (\alpha,\beta)=
\frac{R^{++}_\mathrm{ab}-R^{+-}_\mathrm{ab}-R^{-+}_\mathrm{ab}+R^{--}_\mathrm{ab}}
     {R^{++}_\mathrm{ab}+R^{+-}_\mathrm{ab}+R^{-+}_\mathrm{ab}+R^{--}_\mathrm{ab}}.
\end{equation}
Because of this normalization, it is quite insensitive to fluctuations in the total rate of detected pairs. Reducing the number of detected pairs decreases the statistical accuracy of the measured correlation, but it does not change this correlation per se. It can be compensated simply by increasing the power of the pump accordingly, and by increasing the acquisition time, which we have done in some of the experimental runs reported below when the rate of detected pairs was too low.

A direct way to assess and fine-tune the quality of the produced polarization-entanglement is to measure the correlation when Alice and Bob have their settings set at diagonal in polarization space, which corresponds to $\alpha=\beta=\pi/8$ for the half-wave plates located in front of their respective PBS. Indeed, it is in the diagonal bases that the visibility of the correlation is naturally the lowest, especially when the photons become distinguishable by some degree of freedom other than polarization. Any departure from the optimal conditions reduces the absolute value of the correlation in the diagonal bases, whereas it is much more insensitive to imperfections in the horizontal or vertical basis, as even distinguishable photons $|\mathrm{H}\rangle$ and $|\mathrm{V}\rangle$ lead to a good correlation in these bases. We have therefore measured the correlation in the diagonal bases while varying the temperature of the ppKTP crystal. The result is displayed in Fig.~\ref{VaryTemp}-b.

At optimum temperature ($35.1^{\circ}\mathrm{C}$), the correlation in the diagonal bases $\alpha=\beta=\pi/8$ is close to -1, as expected from the singlet state $|\psi^-_\mathrm{HV}\rangle$ of Eqs.~(\ref{singlet}). For small temperature variation of the ppKTP crystal away from the optimal temperature, the absolute value of the correlation in the diagonal bases decreases, which could be tempting to interpret as caused by a loss of indistinguishability between the photons, as the the center of the spectral distribution of the signal and idler photons start to differ more and more. However, the surprising feature revealed in Fig.~\ref{VaryTemp}-b is that when departing further away from the optimum temperature, the correlation does not in any way remain close to zero---as would be expected from distinguishable $|\mathrm{H}\rangle$ and $|\mathrm{V}\rangle$ photons observed in the diagonal bases---but oscillates instead with increasing amplitude, until the correlation reaches again absolute values close to unity. It should be noted that these experimental features depends strongly on the use of the 1 nm bandwidth interference filter \footnote{We have observed that using instead an interference filter with 10 nm bandwidth, the minima and maxima observed in Fig.~\ref{VaryTemp}-b had their amplitudes greatly reduced, except at the optimum temperature where the correlation was still quite close to -1. It would seem to indicate that it is crucial that the photons remain indistinguishable by their wavelength in order to observe the features of Fig.~\ref{VaryTemp}-b.}.

The temperatures for which the correlation becomes positive and close to 1 are particularly interesting. It happens below the optimum temperature at 28.6$^{\circ}\mathrm{C}$, 25.0$^{\circ}\mathrm{C}$, and $21.8^{\circ}\mathrm{C}$. Indeed, a positive correlation means that the photons measured in the diagonal bases ($\alpha=\beta=\pi/8$) share the same polarization, which is surprising given that the downconverted photons are of type-II, that is, orthogonally polarized. We actually have observed that the orthogonality can actually still be seen quite clearly when Alice's and Bob's fix the orientation of their half-wave plates at $\alpha=\beta=0$, instead of at diagonal. The correlation is then very close to -1 at all temperature. The shift from a -1 correlation in the rectilinear bases to a +1 correlation in the diagonal basis is not at all what one would expect from a singlet state, and therefore needs to be clarified.

\subsection{Extremal points of the diagonal correlation}
\begin{figure}
\begin{center}
a)\includegraphics[width=8.1cm]{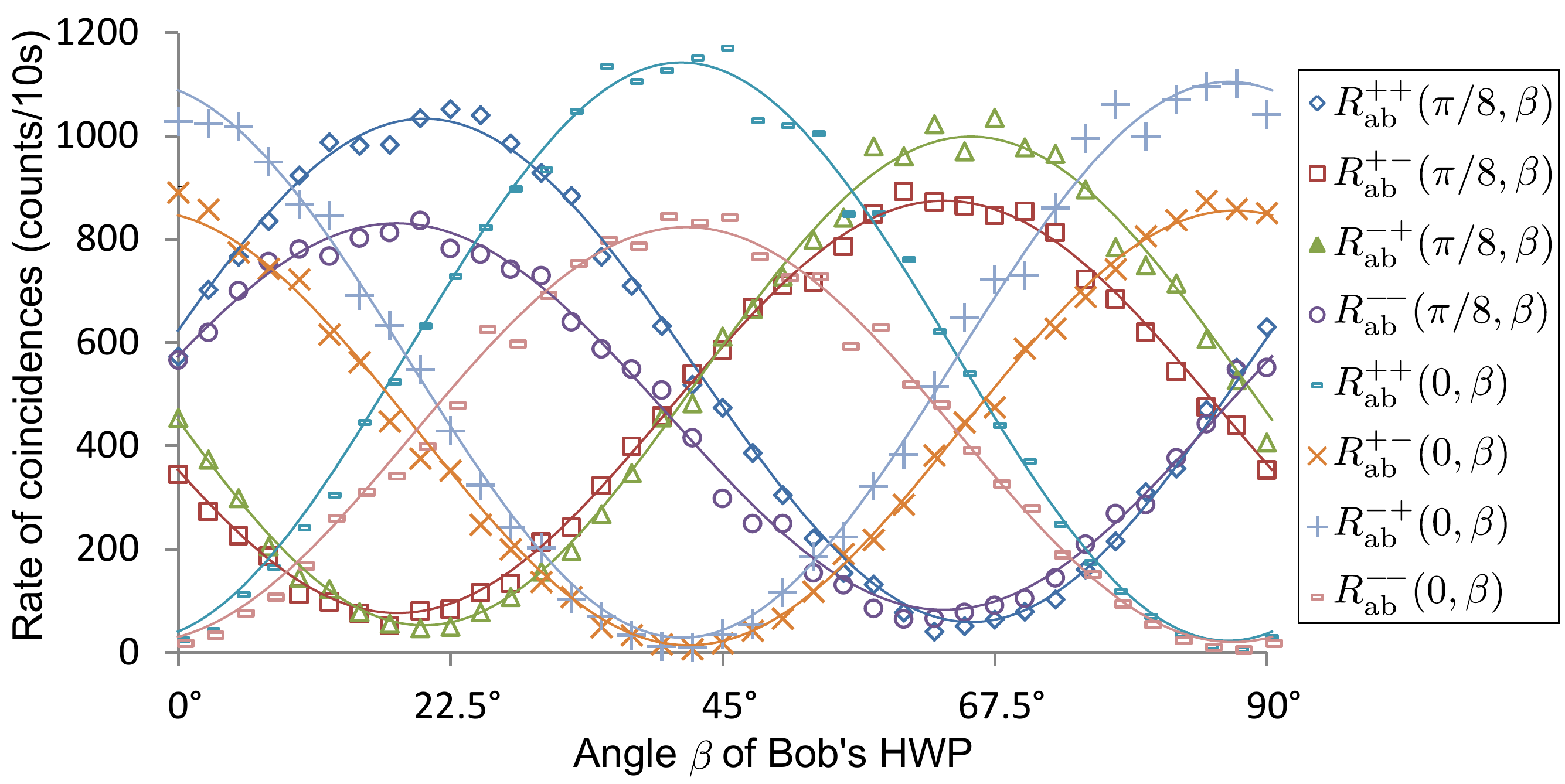}
b)\includegraphics[width=8.1cm]{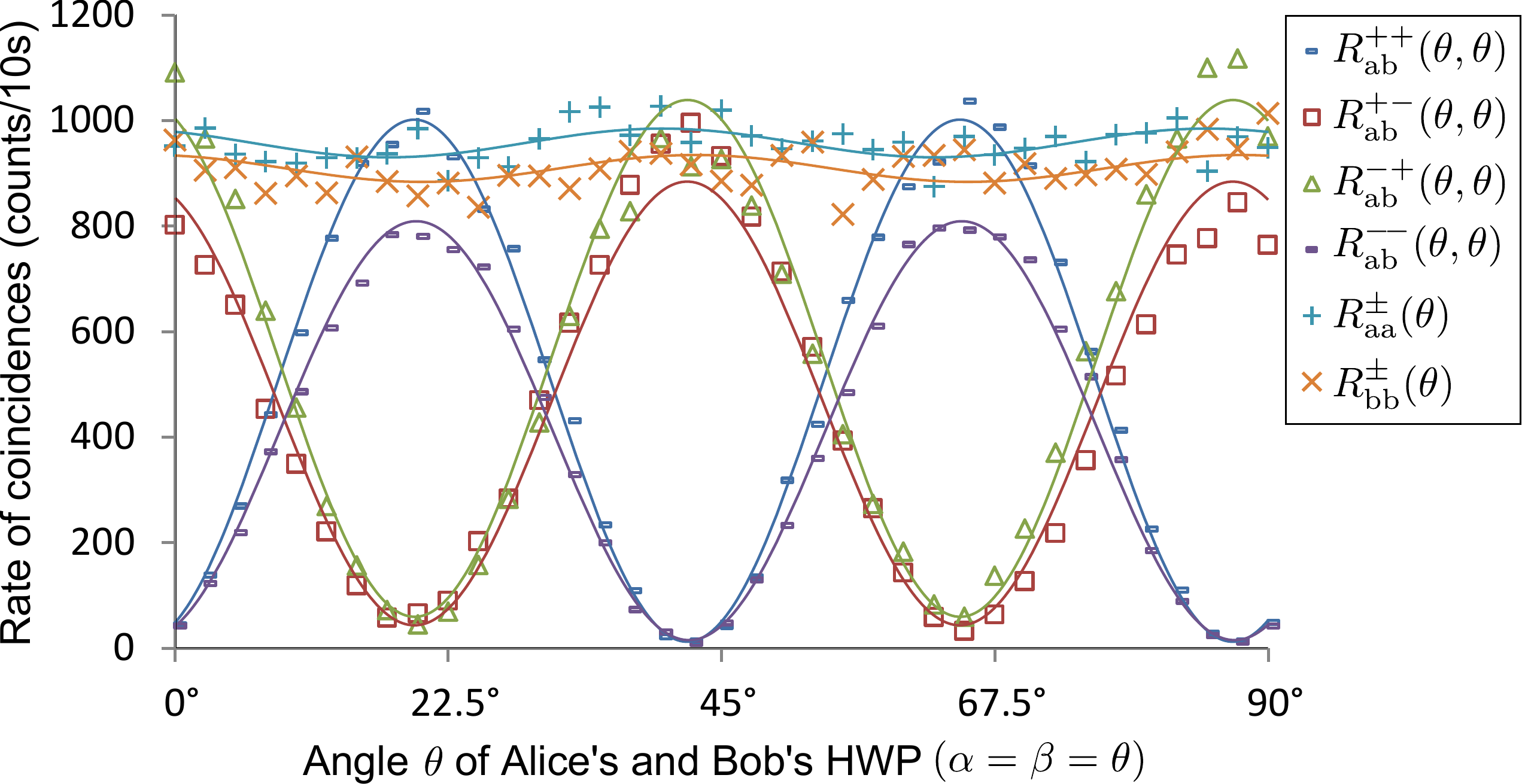}
\end{center}
\caption{\label{FigNonopt} (Color online) a) Rates of coincidences in fixed bases while the temperature of the crystal is maintained at the non-optimum temperature of $25.0^\circ$C, which corresponds to a positive correlation close to 0.9 in Fig.~\ref{VaryTemp}-b. Alice keeps her measurement setting fixed, at $\alpha=0$ (rectilinear basis) or $\alpha=\pi/8$ (diagonal basis), while Bob's HWP setting $\beta$ are varied from 0 to 90. The coincidences exhibit a visibility of 98.2\% in the rectilinear basis and of 87.7\% in the diagonal basis (without substraction of accidental). b) Rates of coincidences and double-counts during a twin scan, while the temperature of the ppKTP crystal is maintained at the non-optimum temperature of $25.0^\circ$C. The results are very much the opposite of what we observed at optimum temperature in Fig.~\ref{FigOpt}-b: the coincidences between Alice and Bob exhibit a strong angular dependence with $\alpha=\beta=\theta$, whereas the doubles are nearly rotationally invariant.}
\end{figure}
At optimum temperature, as well as at the other temperatures for which $\cos\phi$ is close to -1, the observed coincidences take the form Eqs.~(\ref{expoptC}), albeit with a lower visibility in the diagonal bases that depends on how close to -1 is $\cos\phi$. The state of the photons pairs can then be described with good approximation by the singlet state of Eq.~(\ref{singlet}) for the coincidences.

The cases for which the correlation in the diagonal basis seen in Fig.~\ref{VaryTemp}-b is close to +1 is more problematic. This happens for instance when the ppKTP crystal temperature is maintained at $25.0^\circ$C. We have measured the coincidences during fixed scans at this specific temperature (see Fig.~\ref{FigNonopt}-a). The observed rates of coincidences then take the form:
\begin{equation}\label{expnonoptfermC}
\begin{aligned}
    R^{++}_\mathrm{ab} \approx R^{--}_\mathrm{ab}  &\varpropto \frac{1}{2} \sin^2 2(\alpha+\beta)\\
    R^{+-}_\mathrm{ab} \approx R^{-+}_\mathrm{ab}  &\varpropto \frac{1}{2} \cos^2 2(\alpha+\beta).
\end{aligned}
\end{equation}
They exhibit a visibility of 98.2\% in the rectilinear basis and of 87.7\% in the diagonal basis, which is less than in the optimal conditions, but sufficient to clearly violate a Bell inequality under the assumption of fair sampling (as it amounts to a CHSH function $S\simeq2.63$), thereby indicating polarization entanglement. The angular dependence of the coincidences in Eqs.~(\ref{expnonoptfermC}) is however with the sum $\alpha+\beta$, instead of the difference $\alpha-\beta$ expected for a singlet state and observed in Fig.~\ref{FigOpt}-a and Eqs.~(\ref{expoptC}).

The nature of the polarization state can be seen as well during a twin scan, during which Alice and Bob vary their measurement settings while keeping $\alpha=\beta$ at all time. The coincidences between Alice and Bob then exhibit a modulation with high visibility, whereas the double-counts are nearly rotationally invariant (see Fig.~\ref{FigNonopt}-b), which is the opposite behavior of what we had at optimum temperature (see Fig.~\ref{FigOpt}-b). The behavior of the double-counts is particularly striking. They are as numerous as the coincidences between Alice and Bob, and they are nearly rotationally invariant. It suggests that the photons at the output of the polarizing beam-splitters are always exiting through \emph{different} output ports, regardless of the orientation of the HWP in front of the PBS, in a typically fermionic fashion.

\section{Bosonic coalescence and fermionic anti-coalescence}

\begin{figure}
\begin{center}
\includegraphics[width=8cm]{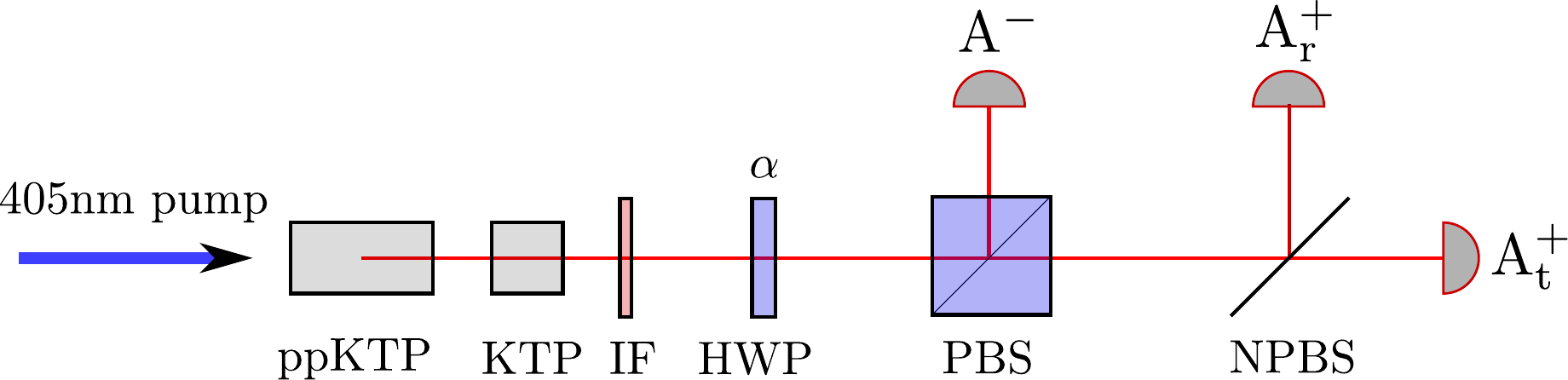}
\end{center}
\caption{\label{figFerm} (Color online) Simple experimental setup to illustrate the fermionic behavior of our photons. The source of down converted photons is the same as previously. Only the analyzing part is changed. The photons are being sent directly to polarization measuring station: a half-wave plate (HWP) rotates the polarization of the photons, and a PBS projects them to a fixed basis. In the transmitted channel at the output of the PBS, a non polarizing beam splitter with two detectors is placed in order to detected the cases when the two down converted are both transmitted at the PBS.}
\end{figure}

In order to investigate further the bosonic and fermionic behavior of our photons, we have performed a simple experiment meant to accommodate for the absence of photon-number resolving capability of our detectors. The purpose of this new experimental setup (see Fig.~\ref{figFerm}) is to monitor how many photons are allowed simultaneously in each output port of a PBS. The simple question that we want to ask with this setup is the following: can the downconverted photons share the same state at the output of the PBS?

In order to give an experimental answer to this question, we fix the temperature of the ppKTP crystal to a specific value, and we vary the orientation of the HWP located in front of the PBS.

\subsection{Experimental setup}
We use the same source of collinear SPDC photons as described in section \ref{Expsetup}. The only difference is in the analyzing part, which is designed to analyze further the properties of the double-counts obtained in the previous setup. The collinear SPDC photons are therefore sent directly on the same input port of Alice's PBS. The reflected output of the PBS (vertically polarized) feeds a detector labelled $\mathrm{A^-}$, as previously. In the transmitted output of the PBS (horizontally polarized), a 50/50 NPBS splits the signal and feeds two detectors labelled $\mathrm{A^+_t}$ and $\mathrm{A^+_r}$. The purpose of this is simply to detect the occurrence of cases in which \emph{both} photons share the same output state $|\mathrm{H}\rangle$, thus exiting together in the transmitted output of the PBS. When it happens, each photon then has a $50\%$ chance of being dispatched independently to either output of the NPBS located after the PBS, and can therefore trigger a coincidence between detectors $\mathrm{A^+_t}$ and $\mathrm{A^+_r}$. We label this rate coincidences as $R^{2,0}_\mathrm{tr}$, to indicate that the downconverted photons were \emph{both} detected in the transmitted output of the PBS, and none in the reflected output.

We also monitor the rate of coincidences between $\mathrm{A^-}$ and $\mathrm{A^+_t}$, as well as the coincidences  between $\mathrm{A^-}$ and $\mathrm{A^+_r}$, which we label respectively $R^{1,1}_\mathrm{t}$ and $R^{1,1}_\mathrm{r}$, to indicate that one photon was detected in the reflected output of the PBS and another one was detected in the transmitted output of the PBS (either in $\mathrm{A^+_t}$ or in $\mathrm{A^+_r}$). The rates $R^{1,1}_\mathrm{t}$ and $R^{1,1}_\mathrm{r}$ thus correspond to the cases when the photons exit the PBS through \emph{different} output.

\subsection{Experimental results}

\begin{figure}
\begin{center}
a)\includegraphics[width=8.1cm]{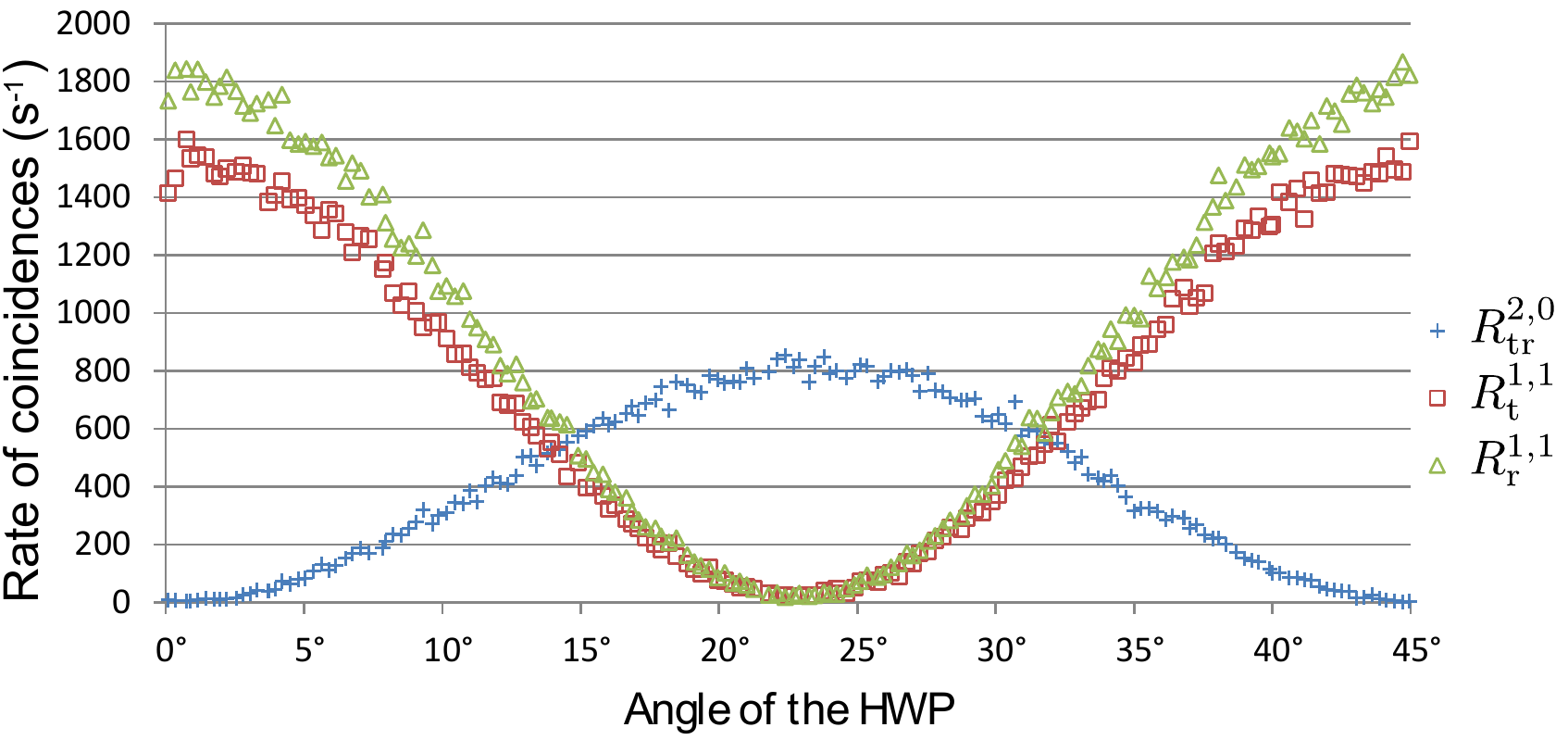}
b)\includegraphics[width=8.1cm]{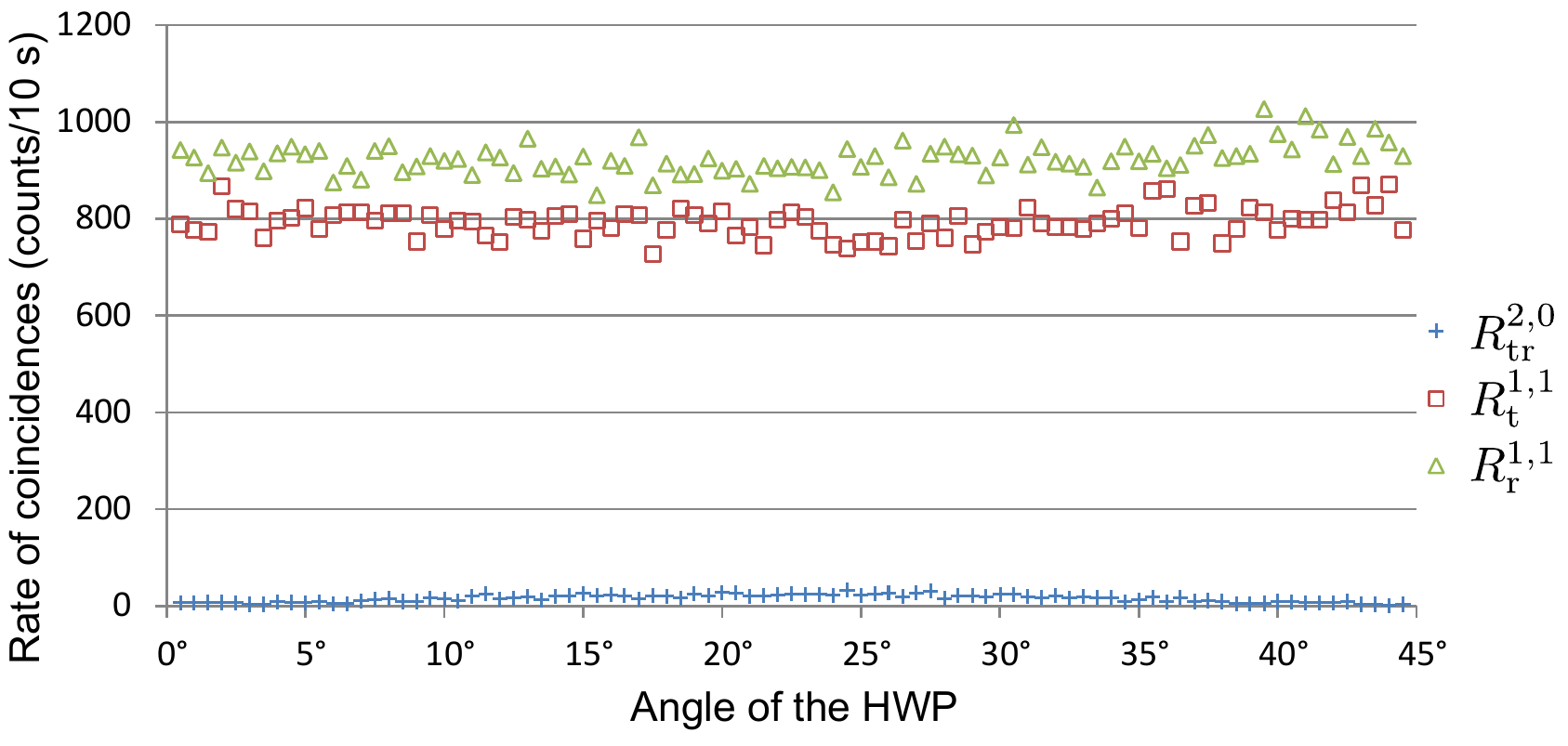}
\end{center}
\caption{\label{Coalescence} (Color online) a) At optimum temperature, the photons exhibit a bosonic behavior. They can exit through the same output, sharing the same polarization and spatial mode, as can be seen from the significantly high rate $R^{2,0}_\mathrm{tr}$, in particular in the diagonal basis when the photons are indistinguishable by all degrees of freedom. The photons then always exit the PBS through the same output port. b) At non-optimum temperature, the photons exhibit a fermionic behavior. The photons refuse to share the same output state at the PBS, and always exit through different port, as can be seen from the nearly rotationally invariant large rate of $R^{1,1}_\mathrm{t}$ and $R^{1,1}_\mathrm{r}$, and near absence of $R^{2,0}_\mathrm{tr}$ rates.}
\end{figure}
\subsubsection{Optimal temperature}

The experimental result obtained with this new setup at optimal temperature is displayed in Fig.~\ref{Coalescence}-a.

When the HWP in front of the PBS is set at 0 or $45^\circ$ (rectilinear basis), the photons are distinguishable by their orthogonal polarizations $|\mathrm{H}\rangle$ and $|\mathrm{V}\rangle$ and therefore exit the PBS through different output with certainty: the horizontal photon $|\mathrm{H}\rangle$ is transmitted, and the vertical photon $|\mathrm{V}\rangle$ is reflected. A large number of $R^{1,1}_\mathrm{t}$ and $R^{1,1}_\mathrm{r}$ is observed; with nearly no coincidences $R^{2,0}_\mathrm{tr}$.

In the diagonal basis (at $\alpha=22.5^\circ$), the photons are no longer distinguishable by their polarization, so that they are indistinguishable by all degrees of freedom, and they coalesce by virtue of their being bosonic \cite{Nam03}. The only rate of coincidences that remains is $R^{2,0}_\mathrm{tr}$, which shows that the photons always exit the PBS through the \emph{same} output port, in a typical bosonic behavior. The photons can share the same quantum state without problem, as bosons are expected to. In fact, the probability $P(2,0)$ that the two photons occupy the same quantum state is enhanced to nearly unity compared to what one would expect from mere combinatorial of distinguishable and independent particles having each a 50$\%$ chance to be transmitted or reflected at the PBS. At optimal temperature, the photons are coalescing when indistinguishable.

\subsubsection{Non-optimal temperature ($25.0^\circ$C)}

The experimental result obtained at the non-optimal temperature of $25.0^\circ$C is displayed in Fig.~\ref{Coalescence}-b.

In the rectilinear basis (0 or $45^\circ$), the behavior is the same as at optimum temperature, and can be understood in the same way: the photons are distinguishable by their orthogonal polarizations $|\mathrm{H}\rangle$ and $|\mathrm{V}\rangle$. They therefore exit the PBS through different output with certainty: the horizontal photon $|\mathrm{H}\rangle$ is transmitted, and the vertical photon $|\mathrm{V}\rangle$ is reflected. A large number of $R^{1,1}_\mathrm{t}$ and $R^{1,1}_\mathrm{r}$ is observed; with nearly no coincidences $R^{2,0}_\mathrm{tr}$.

However, what is striking is that this behavior remains regardless of the orientation of the HWP. Even in the diagonal basis (at $\alpha=22.5^\circ$), the rate of coincidences $R^{2,0}_\mathrm{tr}$ between the detectors located in the transmitted output of the PBS remains minimum, which shows that the photons refuse to share the same spatial and polarization mode at the output of the PBS, regardless of the setting of the HWP in front of the PBS, and even when they are indistinguishable by all degrees of freedom. The photons are anti-coalescing, even when indistinguishable, and always exit through different port.

Note that the absence of $R^{2,0}_\mathrm{tr}$ coincidences cannot be explained by assuming that the photons do exit through the same output port of the PBS, and behave in the same coalescing way at the output of the NPBS. If this was so, it would certainly explain the absence of $R^{2,0}_\mathrm{tr}$ coincidences, but it would also produce no $R^{1,1}_\mathrm{t}$ and $R^{1,1}_\mathrm{r}$ coincidences.  It is the combined observation of the large number of $R^{1,1}_\mathrm{t}$ and $R^{1,1}_\mathrm{r}$ coincidences with the near absence of $R^{2,0}_\mathrm{tr}$ coincidences that indicates clearly the anti-coalescent behavior, which is characteristic of fermions. We have observed the same behavior when replacing the HWP by a quarter-wave plate (QWP), which explores a different trajectory in the Poincar\'{e} sphere as the setting of the QWP is changed. This anti-coalescent behavior remains, as one would expect from fermions obeying the Pauli exclusion principle.

\section{Conclusion}

It should be clear that the standard theoretical description \cite{Rubin94,TSuhara,TSuhara2007,Martin08,Martin10,KuklewiczPhD,Sciarrino} that we used in Section \ref{theory} is insufficient to account for these observations. In particular, the fermionic anti-coalescence observed at non-optimum temperature would only be explainable by a singlet state in the polarization degree of freedom \emph{before} the NPBS, when the photons are still collinear. It is indeed notorious that the singlet state is rotationally invariant (and antisymmetric), which would explain the persistent anti-coalescent behavior regardless of the use of a HWP or of a QWP. However, with an identical spatial mode for both photons, this would mean that the global quantum state is antisymmetric, which is of course problematic for pairs of photons. It would be possible to recover symmetry with the use of an additional degree of freedom that would also be entangled and antisymmetric, but the nature of this additional degree of freedom and the theoretical account of its antisymmetric form for pairs of photons remain to be justified, and will in fact be the subject of a forthcoming publication \cite{Adenier13}.

This work was supported by a special grant from the Department of Physics and The Faculty of Mathematics and Natural Sciences at the University of Oslo.

\end{document}